\documentstyle[prl, aps, multicol]{revtex}
\input{psfig}

\def\be{\begin{equation}}
\def\ee{\end{equation}}
\def\ba{\begin{eqnarray}}
\def\ea{\end{eqnarray}}

\begin{document}
\draft
\title{Magnetic breakdown in a normal-metal--superconductor\\ 
proximity sandwich}

\author{Alban L.\ Fauch\`ere and Gianni Blatter}

\address{Theoretische Physik, Eidgen\"ossische Technische Hochschule,
 CH-8093 Z\"urich, Switzerland }

\date{\today}

\maketitle

\begin{abstract}
We study the magnetic response of a clean normal-metal slab of finite 
thickness in proximity with a bulk superconductor. We determine its free
energy and identify two (meta-)stable states, a diamagnetic one where the 
applied field is effectively screened, and a second state, where the field
penetrates the normal-metal layer. We present a complete characterization of 
the first order transition between the two states which occurs at the 
breakdown field $H_b\left(T\right)$, including its spinodals, the jump in the 
magnetization, and the latent heat. The bistable regime terminates 
at a critical temperature $T_{crit}$ above which the sharp transition is 
replaced by a continuous cross-over. We compare the theory with experiments 
on normal-superconducting cylinders.
\end{abstract}

\begin{multicols}{2}

The superconducting proximity effect in a normal metal adjacent to a
superconductor has received a revived interest in the past 
decade\cite{proxi}. Among the fundamental equilibrium problems, the magnetic 
response of normal-metal--superconductor (NS) structures deserves particular 
interest. Experiments have demonstrated the non-trivial 
screening properties of these hybrid structures, exhibiting a hysteretic 
magnetic breakdown at finite fields\cite{oda,mota1,bergmann} as well as a 
presently unexplained re-entrance in the magnetic susceptibility at low 
temperatures\cite{motaprl}. The investigated samples have 
typical dimensions comparable to the coherence length $\xi_N$ of the 
normal metal, 
attributing a key role to the quantum coherence of the electrons coupled to 
the macroscopic phase of the superconductor.

\begin{figure}
\noindent
\centerline{\psfig{figure=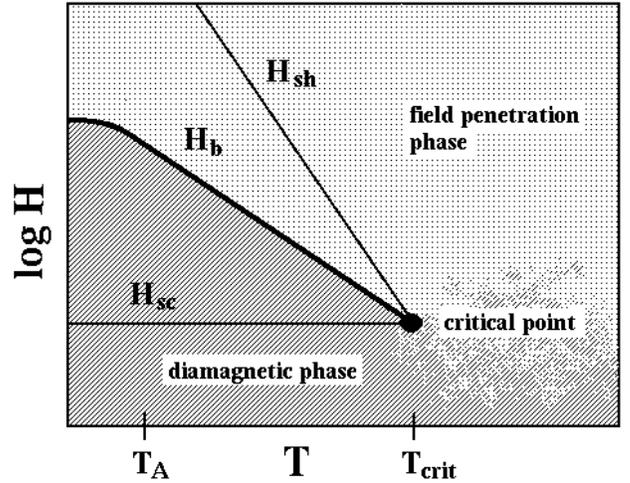,width=85mm}}
\narrowtext
\caption[*]{$H-T$ phase diagram of the normal metal slab of thickness $d$. The 
breakdown field $H_b\left(T\right)\sim \Phi_{\circ}/\lambda_N\left(T\right)d$ 
marks the first order transition between the 
diamagnetic and the field penetration phase ($\lambda_N\left(T\right)$ denotes 
the penetration depth, $\Phi_{\circ}$ the superconducting flux unit). 
The critical point at the intersection of the spinodals 
$H_{sc}\left(T\right)\sim \Phi_{\circ}/d^2$ and 
$H_{sh}\left(T\right)\sim \Phi_{\circ}/\lambda_N^2\left(T\right)$  
separates the first order transition for $\lambda_N\left(T\right)\ll d$ 
from the continuous cross-over at large temperature 
($\lambda_N\left(T\right)\gg d$).}
\label{fig1}
\end{figure}

The self-consistent study of the screening currents in a NS sandwich 
within the framework of the Ginsburg-Landau (GL) equation
was carried out a long time ago by the Orsay group\cite{orsay}. 
Their work has provided the 
first understanding of the non-linear field phenomena such as the magnetic 
breakdown. However, in the proximity effect, the GL equations are at their 
limit of validity and their use is restricted to the dirty limit. 
The quasi-classical Green's function 
technique\cite{shelankov} allows to describe the clean limit 
using the Eilenberger equations\cite{eilen} and to generalize the dirty limit 
results using the Usadel equations\cite{usadel}. Zaikin was the first to derive
the magnetic response of a normal-metal slab of finite thickness connected to 
a bulk superconductor along these lines\cite{zaikin}. Most notably, he found a 
non-local screening behavior in the clean limit linear response which has 
an appealing similarity to the one found in superconductors of the Pippard 
type. The applied (static) magnetic field was found to be over-screened, the 
magnetic induction changing sign inside the normal layer. 
Using numerical methods, Belzig {\sl et al.}\cite{belzig} have 
investigated the non-linear 
field regime of these equations and found two (meta-)stable mean-field solutions 
in both the clean and the dirty limit. 
In this work, we determine the $H$-$T$ 'phase' diagram shown in Fig.\ 
\ref{fig1} of the normal metal layer in the clean limit, where the bistable 
regime is particularly extended. In thermodynamic 
equilibrium, we find a magnetic breakdown at $H_b\left(T\right)$, which 
is a first order transition separating the phase of diamagnetic screening from 
the phase of magnetic field penetration. 

Mota {\sl et al.}\cite{motapd,motab} have recently investigated the magnetic
response of metallic cylinders with a superconducting core. Their data, which
is not described by the results of the GL equations, was claimed to be 
characteristic for the ballistic limit\cite{motab}.
This has motivated us to derive the analytic dependence of the clean limit 
expression for the breakdown field $H_b$ on temperature $T$ and thickness $d$ 
of the normal layer and to compare it to the experiment. From the free energy
of the normal layer, which allows us to identify the two (meta-)stable states, 
we determine the spinodals, the 
thermodynamic breakdown field $H_b\left(T\right)$ and find the 
jumps in magnetization and entropy at the transition. 
Furthermore, we obtain a critical temperature which marks the upper limit of 
the bistable regime (see Fig.\ \ref{fig1}). Finally, we work out the signatures
of the non-locality in the ballistic regime as they show up in the magnetic 
susceptibility $\chi$ and compare them with the 
experimental data. The following discussion is divided into four sections, 
the analysis of the constitutive relations (Sec.\ I), the solution of the 
magnetostatic problem (Sec.\ II), the determination of the breakdown field 
from the free energy (Sec.\ III), and finally the comparison with experiment
(Sec.\ IV).

\section{Constitutive relations}
 
The quasi-classical Green's function technique provides an appropriate 
description of a metal with nearly spherical Fermi surface. In a finite 
magnetic field, the vector potential 
${\bf A}\left({\bf x}\right)$ can be included as a phase factor along 
unperturbed trajectories, provided the dimensions of the normal metal are 
smaller than the Larmor radius ($r_L=p_Fc/eH$, cyclotron radius of an 
electron traveling at Fermi velocity). In the ballistic limit, the 
quasi-classical $2\times 2$ matrix Green's function 
$\hat{g}_{\omega_n}\left({\bf x},{\bf v}_F\right)$ satisfies the Eilenberger 
equation\cite{eilen} 
($e=|e|$, $\hbar=k_B=c=1$), 
\ba && -\left({\bf v}_F\cdot {\bf \nabla}\right) 
\hat{g}_{\omega_n}\left({\bf x},{\bf v}_F\right)= \nonumber \\ &&
\left[ \left(\omega_n+ie {\bf v}_F\cdot {\bf A}\left({\bf x}\right) \right) 
\hat{\tau}_3  + \Delta\left({\bf x}\right) \hat{\tau}_1, 
\hat{g}_{\omega_n}\left({\bf x},{\bf v}_F\right) \right], \label{eilenberger}
\ea 
where the mean-field order parameter $\Delta$ provides an off-diagonal 
potential ($\hat{\tau}_i$ denote the Pauli matrices, $\omega_n=\left(
2n +1\right)\pi T$ are the Matsubara frequencies, $\left[\cdot,\cdot\right]$ 
is the commutator). We have excluded elastic scattering processes by 
assuming $T\gg 1/\tau_{el}$. 

We consider a  normal metal slab of thickness $d$ 
on top of a bulk superconductor as shown in the inset of Fig.\ \ref{fig1}. 
The vector potential ${\bf A}=\left(0,A\left(x\right),0\right)$ describes a 
magnetic field ${\bf B}= \left(0,0,B\left(x\right)\right)$ applied parallel to 
the surface, which induces screening currents 
${\bf j}=\left(0,j\left(x\right),0\right)$. We make the following 
idealizations in the description of the NS sandwich: The superconducting order 
parameter follows a step function
$\Delta\left(x\right)=\Delta \theta\left(-x\right)$ ($\Delta$ real), no 
attractive interactions being present in the normal layer. We assume a perfect 
NS interface as well as specular reflection at the normal-vacuum 
boundary. 

In the subsequent analysis we restrict our attention to the magnetic response 
of the normal layer. In the proximity effect, the macroscopic coherence of the 
superconducting condensate induces correlated electron-hole pairs in the 
normal layer through the process of Andreev reflection. The basic process 
consists of
an electron traveling forward and a hole traveling backward along a 
quasi-classical trajectory as shown in Fig.\ \ref{fig2} (at discrete energies, 
bound Andreev states are found along these trajectories).
In the presence of a magnetic 
field, the area enclosed by the trajectory (see Fig.\ \ref{fig2}) is threaded 
by the flux 
\be \Phi\left(a,\vartheta,\varphi\right)= \oint {\bf A}\left({\bf x}\right) 
\cdot d{\bf x} = 2\tan\vartheta \cos\varphi \int_0^d A\left(x\right) dx,
\ee
which can be expressed through the integral $a=\int_0^d A\left(x\right) dx$ 
times a geometric factor due to the inclination of the trajectory (the 
spherical angles $\vartheta$ and $\varphi$ parameterize the direction of the 
trajectory). 
The current carried along a trajectory depends on the phase factor 
$2\Phi\left(a,\vartheta,\varphi\right)/\Phi_{\circ}$ acquired by the 
propagation 
of both the electron and the hole along the Andreev loop, and we arrive 
at an intrinsic non-local current--field dependence $j\left(a\right)$.
The total current is determined by the sum over the currents along the 
quasi-classical trajectories,
\be j_y\left(x\right) = \frac{iemp_F}{\pi} T\sum_{\omega_n>0} 
\langle v_y \mbox{Tr}\left[\hat{\tau}_3 
\hat{g}_{\omega_n}\left(x,{\bf v}_F\right) \right] \rangle \label{current}
\ee
(the brackets $\langle \mbox{ ... } \rangle$ denote the average over the 
angles $\vartheta$, $\varphi$), from which we reproduce the expression 
first derived by Zaikin\cite{zaikin},
\be j\left(a\right) = \int_0^{\pi/2} d\vartheta \int_0^{\pi/2} d\varphi \, 
j\left( \vartheta, \varphi, \Phi\left(a,\vartheta,\varphi\right)  \right), 
\label{jy} \ee
where ($\alpha_n=2\omega_nd/v_F\cos\vartheta$)
\ba && j\left(\vartheta,\varphi, \Phi\left(a,\vartheta,\varphi\right) \right)
=-\frac{2e p_F}{\pi^2} T\sum_{\omega_n>0} \sin^2\vartheta \cos\varphi 
\label{jgeneral} \\ && \times
\frac{\Delta^2 \sin 2\pi\Phi/\Phi_{\circ}}{\left(\omega_n \cosh\alpha_n 
+ \sqrt{\omega_n^2+\Delta^2}\sinh\alpha_n\right)^2
+\Delta^2\cos^2\pi\Phi/\Phi_{\circ}}. \nonumber 
\ea
Note that $j$ is independent of $x$. The induced currents 
for each trajectory depend only on the flux $\Phi$ modulo the 
superconducting flux quantum $\Phi_{\circ}=\pi\hbar c/e$, reflecting
gauge invariance. At small fields ($a/\Phi_{\circ}\ll 1$),
the current response is diamagnetic for all trajectories and the proximity 
effect produces screening currents in the normal metal. 
As the field increases to $a/\Phi_{\circ} \sim 1$, some of the more extended 
trajectories produce paramagnetic currents, since the reduced flux 
$\Phi \in \left[-\Phi_{\circ}/2,\Phi_{\circ}/2\right]$ they enclose becomes 
negative, and the net diamagnetic current response is reduced. 
As we reach large fields ($a/\Phi_{\circ} \gg 1$), the Andreev levels become 
mutually dephased due to a uniform distribution of the reduced flux. The 
associated currents are randomly
dia- or paramagnetic and the net current vanishes. Note that the proximity 
effect, i.e., the existence of the Andreev levels is not 
destroyed in this limit, leading to a finite kinetic energy of the 
currents induced by the magnetic field. 

\section{Magnetostatics}
 
Owing to the independence of $j$ on $x$, the Maxwell equation 
$-\partial_x^2A\left(x\right)=4\pi j$ and the constitutive equation (\ref{jy}) 
combined with the boundary conditions\cite{bound} $A\left(x=0\right)=0$ and 
$\partial_xA\left(x=d\right)=H$ can be given a formal solution. 
We arrive at a parabolic dependence for $A\left(x\right) = 
Hx + 4\pi j\left(a\right) x\left(d-\frac{x}{2}\right)$ parameterized by 
$a=\int_0^d A\left(x\right) dx$, which in turn is determined through the 
self-consistency condition
\be  a=\frac{Hd^2}{2} + \frac{4\pi}{3} j\left(a\right) d^3. 
\label{selfconsistency} 
\ee
The total magnetization $\cal M$ (per unit surface) is defined by 
\be 
4\pi{\cal M}= \int_0^d dx\, \left(\partial_xA\left(x\right)-H\right) 
= 2\pi j\left(a\right) d^2. \label{magnet}
\ee
Eq.\ (\ref{selfconsistency}) contains the essential physics of the
problem: For small fields ($a\to 0$), the current $j\approx -3H/8\pi d$ 
linearly suppresses the magnetic induction on the geometric length scale 
$d$ to the value $B\left(0\right)\to -H/2$ at the NS boundary. 
The magnetic induction is thus 
over-screened and assumes an opposite sign at the NS interface. A closer look 
shows that the current is given by the linear response expression
\be j\left(a/\Phi_{\circ}\ll 1\right) \approx 
-\frac{1}{4\pi\lambda_N^2\left(T\right)d} a, \label{jlinear} 
\ee
which depends on penetration depth $\lambda_N\left(T\right) \ll d$ to be
derived below (Eq. (\ref{lambda})). When inserted back into 
(\ref{selfconsistency}), 
the vector potential is found to be strongly suppressed to 
$a \approx 3 H \lambda^2_N\left(T\right)/2$, and we obtain a consistent 
diamagnetic solution (i.e., $a/\Phi_{\circ}\ll 1$) for fields up to 
$H < \Phi_{\circ}/\lambda^2_N$. 
At large fields, the current vanishes ($j\to 0$) 
and the magnetic field penetrates the normal layer. From Eq.\ 
(\ref{selfconsistency}) we find $a\approx Hd^2/2$, consequently this metallic 
behavior is expected down to magnetic fields $H > \Phi_{\circ}/d^2$, as 
follows from the condition $a/\Phi_{\circ} \gg 1$ for the Andreev levels 
to be dephased.
With $\Phi_{\circ}/d^2 \ll \Phi_{\circ}/\lambda^2_N$ the diamagnetic and 
field penetration solution coexist in the regime $\Phi_{\circ}/d^2 < H < 
 \Phi_{\circ}/\lambda^2_N$. These simple estimates for the limits of the 
bistable regime elucidate the numerical data of Belzig 
{\sl et al.}\cite{belzig}.

\section{Thermodynamics} 

In the phase diagram of Fig.\ \ref{fig1} the upper and lower bounds of the 
bistable regime found from the above mean-field analysis are identified with 
the spinodals of the transition, the super-cooled field 
$H_{sc}\sim \Phi_o/d^2$ and the super-heated field 
$H_{sh} \sim \Phi_o/\lambda_N^2\left(T\right)$. In the
thermodynamic equilibrium, a magnetic breakdown occurs at an intermediate 
field, connecting the diamagnetic regime to the field penetration regime by a 
first order transition. In the following, we determine this breakdown field 
and the associated entropy and magnetization jump from the free energy.

The energy (per unit surface) of the currents $j\left(x\right) = 
-\delta F/\delta A\left(x\right)$ is obtained via an integration over the 
non-linear current expression, 
\ba F\left(a\right)&=& -\int_0^a j\left(a'\right) da' \nonumber \\
&=& \frac{p_F^2}{\pi^2}T\sum_{\omega_n>0}\int_0^{\pi/2} d\vartheta 
\int_0^{\pi/2} d\varphi \sin\vartheta \cos\vartheta \label{fofa} \\
&& \!\!\!\!\!\! \!\!\!\!\!\!\!\!\!\!\!\!\! \log\frac{\left(\omega_n\cosh
\alpha_n+\sqrt{\omega_n^2+\Delta^2}\sinh\alpha_n\right)^2 +\Delta^2}
{\left(\omega_n\cosh\alpha_n+\sqrt{\omega_n^2+\Delta^2}\sinh\alpha_n\right)^2 
+ \Delta^2\cos^2\pi\Phi/\Phi_{\circ}}. \!\!\!\! \nonumber
\ea
$F\left(a\right)$ describes the difference in free energy between the metal 
layer under proximity and in the normal state. $F\left(a\right)$ is a 
monotonous and strictly positive function, reflecting the absence of 
condensation energy in the normal layer, and expresses the cost of the 
induced proximity effect lying in the kinetic energy of the currents induced 
by the vector potential. 
The free energy ${\cal F}\left(T,{\cal M}\right)$ is constructed by adding the 
electro-magnetic field energy and subtracting the vacuum field 
contribution, 
\be {\cal F}\left(T,{\cal M}\right)= F\left(a\right) + \int_0^d dx\,
\left(\frac{\left(\partial_x A\left(x\right)\right)^2}{8\pi}-\frac{H^2}{8\pi}
\right) . \label{free}
\ee
We do not include the condensation energy and the kinetic energy of the 
screening currents in the superconductor. The field dependent term of the 
condensation energy might in fact be of the order of the free energy in the 
normal 
layer and would be expected to produce numerical corrections in the results, 
which are not accounted for by our idealized choice of the order parameter
$\Delta\left(x\right)=\Delta \theta\left(-x\right)$. The kinetic energy of 
the screening currents $\sim H^2\lambda$ may be neglected.

\begin{figure}[htb]
\noindent
\centerline{\psfig{figure=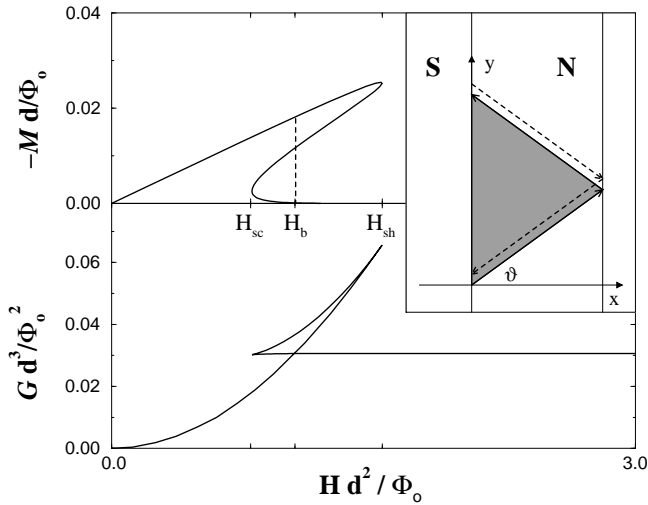,angle=-90,width=82mm}}
\narrowtext
\vspace{2mm}
\caption[*]{Magnetization ${\cal M}\left(H\right)$ and free energy 
${\cal G}\left(T,H\right)$ at a temperature $T=1.5\,T_A$ ($T_A=v_F/2\pi d$). 
The representation is universal
in the thickness $d$. The small and large field branches represent 
(meta-)stable solutions describing the diamagnetic and field penetration
phases, which overlap in the field interval $H_{sc}<H<H_{sh}$. The Maxwell 
construction determines the first order transition between the phases at the 
breakdown field $H_b$ (dashed line). 
Inset: Cross-section of the normal-metal slab in contact with the bulk 
superconductor. The quasi-classical electron-hole trajectory at angles 
$\vartheta$, $\varphi=0$ encloses a flux that enters as a phase factor 
in the wavefunction.}
\label{fig2}
\end{figure}

After a Legendre transformation, we obtain the Gibb's free energy 
\ba {\cal G}\left(T,H\right)&=& F\left(T,{\cal M}\right) - {\cal M}H \nonumber 
\\ &=& F\left(a\right)+\int_0^d dx\, \frac{\left(\partial_x 
A\left(x\right)-H\right)^2} {8\pi}. \label{freeenergy}
\ea
The field term in Eq.\ (\ref{freeenergy}) describes the work necessary to 
expel the magnetic field. The extrema of the free energy $\cal G$ 
with respect to $a$ reproduce the equation of state (\ref{selfconsistency}). 
Fig.\ \ref{fig2} shows the free energy 
${\cal G}\left(H\right)$ as obtained from the parameterization of $\cal G$ 
and $H$ through $a$\cite{belz2}. The breakdown field $H_b\left(T\right)$ is
determined by the intersection of the free energies $\cal G$ of the two
(meta-)stable solutions. We note that this procedure is equivalent to 
the Maxwell construction in the magnetization curve 
${\cal M}=-\partial{\cal G}/\partial H$ of Fig.\ \ref{fig2}.

In the following, we consider the free energy (\ref{fofa}) in the two 
temperature limits $T=0$ and $T_A\ll T \leq \Delta$ and obtain
($T_A=v_F/2\pi d$),
\ba F_{T=0}\left(a\right)&\approx &\frac{p_F^3}{4\pi^3dm} \int_0^{\pi/2} d
\vartheta 
\int_0^{\pi/2} d\varphi \sin\vartheta \cos^2\vartheta \nonumber \\
&& \left\{ \arctan\left[\tan\pi\Phi/\Phi_{\circ} 
\right]\right\}^2, \label{FT0} \\
F_{T\gg T_A}\left(a\right)&\approx & \frac{4p_F^2T}{\pi^2}\gamma^2
\left(T,\Delta\right) 
\int_0^{\pi/2} d\vartheta \int_0^{\pi/2} d\varphi \sin\vartheta \cos\vartheta
\nonumber \\ && \exp\left(-\frac{2T}{T_A \cos\vartheta}\right) 
\sin^2\pi\Phi/\Phi_{\circ}. \label{FT>}
\ea 
The finite value of the superconducting gap $\Delta$ is accounted for by 
the dimensionless parameter
\be \gamma\left(T,\Delta\right)=\Delta/\left(\sqrt{\Delta^2+
\left(\pi T\right)^2}+\pi T\right) < 1.
\ee

The free energies of the two (meta-)stable stable states can be approximated 
by their asymptotic forms in the limits $a\to 0$ and $a\to \infty$. 
In the diamagnetic regime, the expansion in $a/\Phi_{\circ}$ up to quadratic 
order of (\ref{FT0}) and (\ref{FT>}) provide the result 
\be F\left(a\right) \approx   \frac{a^2}{8\pi\lambda_N^2\left(T\right)d}. 
\label{flinear} \ee
Eq.\ (\ref{flinear}) is valid both in the low and high temperature limits 
using the penetration depth $1/\lambda_N^2\left(0\right)\equiv 1/\lambda_N^2 
= \left(4\pi n e^2/m\right)$ at $T=0$ and 
\be \frac{1}{\lambda_N^2\left(T\right)}\approx  \frac{1}
{\lambda_N^2}\gamma^2\left(T,\Delta\right) 
\frac{6T_A}{T} e^{-2T/T_A} \label{lambda}
\ee
for $T\gg T_A$. Note that the derivative $j = -\partial F/\partial a$
applied to Eq.\ (\ref{flinear}) produces the linear response constitutive 
relation of Eq.\ (\ref{jlinear}).
The Gibb's free energy follows from Eqs.\ (\ref{free}) and (\ref{flinear}), 
using the solution of the Maxwell equations, 
\be {\cal G}\left(a\ll \Phi_{\circ}\right)\approx  
\frac{3}{32\pi}H^2d.  \label{f1}
\ee
Eq.\ (\ref{f1}) is dominated by the magnetization work necessary to expel the 
field, which is parametrically larger (by $\left(d/\lambda_N\left(T\right) 
\right)^2$) than the kinetic energy of the currents. 
 
In the field penetration regime we approximate the free energy
by its asymptotic value at $a\to\infty$. In this limit we replace the
strongly oscillating functions of $\Phi$ in (\ref{FT0}) and (\ref{FT>}) by
their average value $\langle\left(\arctan\tan\Phi\right)^2\rangle=\pi^2/12$ 
and $\langle\sin^2\Phi\rangle=1/2$ and obtain
\ba  {\cal G}_{T=0}\left(a\gg \Phi_{\circ} \right)&\approx &
\frac{1}{384\pi}\frac{\Phi_o^2}{\lambda_N^2d}, \nonumber \\
{\cal G}_{T\gg T_A}\left(a \gg \Phi_{\circ}\right)&\approx 
&\frac{3}{16\pi^3} \gamma^2\left(T,\Delta\right)    
\frac{\Phi_o^2} {\lambda_N^2d} e^{-2T/T_A}. \label{f2}
\ea
The magnetization energy vanishes in this limit. The corrections to the
free energy (\ref{f2}) are of relative order $\left(\Phi_{\circ}/a\right)^2$.

The magnetic breakdown field $H_b\left(T\right)$ is determined by the 
intersection of the two asymptotics of the free energy ${\cal G}$ given by 
Eqs. (\ref{f1}) and (\ref{f2}),
\ba H_b\left(T=0\right)&\approx & \frac{1}{6} \frac{\Phi_o}{\lambda_N d}, 
\label{b0}\\
    H_b\left(T\gg T_A\right)&\approx & \frac{\sqrt{2}}{\pi} 
    \gamma\left(T,\Delta\right)  \frac{\Phi_o}{\lambda_N d} 
   e^{-d/\xi_N\left(T\right)}.
\label{breakdown} \ea
We note three important features of this result: The temperature 
dependence is a simple exponential with the exponent 
$d/\xi\left(T\right)=T/T_A$, where $\xi_N\left(T\right)=v_F/2\pi T$ denotes 
the normal metal coherence length. The amplitude of the breakdown field scales 
inversely proportional to the thickness of the normal layer, $H_b \sim 1/d$. 
In the limit $T\to 0$ the magnetic 
breakdown field saturates to a value which is suppressed by the universal 
factor $\pi/6\sqrt{2}\approx 0.37$ as compared to the extrapolation of the 
high temperature result. 

We arrive at the $H-T$ phase diagram shown in Fig.\ \ref{fig1}. 
The first order transition between the diamagnetic and the field penetration 
regime takes place between the spinodals $H_{sc}\sim \Phi_o/d^2 < 
H_b\left(T\right) < H_{sh}\sim \Phi_o/\lambda_N\left(T\right)^2$ which 
delimit the (meta-)stable regime. Their intersection marks the critical 
temperature 
\be T_{crit}\approx T_A \log \left(d/\lambda_N\right), \label{tcrit}
\ee
where 
$\lambda_N\left(T_{crit}\right) \approx d$. Below $T_{crit}$ the penetration 
depth is small, $\lambda_N\left(T_{crit}\right) < d$, and we observe 
a first order transition. Above the critical point $T_{crit}$, 
where $\lambda_N\left(T_{crit}\right) > d$, a continuous and reversible 
cross-over between the diamagnetic and field penetration regime is expected. 
We note that this distinction is similar to the one between Type I and Type II
superconductors with respect to the penetration depth $\lambda$ and the
superconducting coherence length $\xi$.

The latent heat (at $T\gg T_A$) of the transition follows 
from Eqs.\ (\ref{f1}), (\ref{f2}), 
and (\ref{breakdown}) using $S=-\partial{\cal G}/\partial T$,
\be T\Delta S  \approx  \frac{3}{16\pi} \frac{T}{T_A} H_b^2\left(T\right)d, 
\label{latent}
\ee
and is related to the magnetization jump
\be 4\pi \Delta {\cal M} \approx  \frac{3}{4} H_b\left(T\right)d \label{Mjump}
\ee
via the Clausius-Clapeyron equation.

In the derivation of the breakdown field we have used the asymptotic 
expansions of the free energies in $a/\Phi_{\circ}$ and $\Phi_{\circ}/a$, 
respectively. Their quality at the transition point is determined by the 
range of overlap between the diamagnetic and the field penetration regimes 
in Fig.\ \ref{fig2}, which
is governed by the parameter $\lambda_N\left(T\right)/d$. In the diamagnetic
phase, the corrections are of the order of $\left(a/\Phi_{\circ}\right)^2\sim
\left(H_b\lambda^2_N\left(T\right)/\Phi_{\circ}\right)^2\sim
\left(\lambda^2_N\left(T\right)/d\right)^2$, and similarly in the 
field penetration regime. The expansion thus breaks down at 
$\lambda_N\left(T\right)\approx d$, which is the critical point of the 
transition line.
We note that the total magnetization changes from its diamagnetic value 
${\cal M} \sim H_b d$ to the strongly suppressed value 
${\cal M} \sim H_b d \left(\lambda_N\left(T\right)/ d\right)^2$ at the 
transition, reflecting its strong first order character.

\section{Experiment}
  
Mota 
{\sl et al.}\cite{motapd,motab} have measured the breakdown field in Ag-Nb 
cylinders. The clean limit theory valid for $T\gg v_F/{\sl l}_{el}$ may be 
used provided that ${\sl l}_{el}\gg d$, which is claimed to be fulfilled in 
the experiment. In our comparison we neglect the influence of diffusive 
boundary 
scattering or any potential barrier at the NS interface, and ignore the 
difference in geometry, cylindrical for the sample and planar in the 
theoretical model.
\begin{figure}[htb]
\noindent
\centerline{\psfig{figure=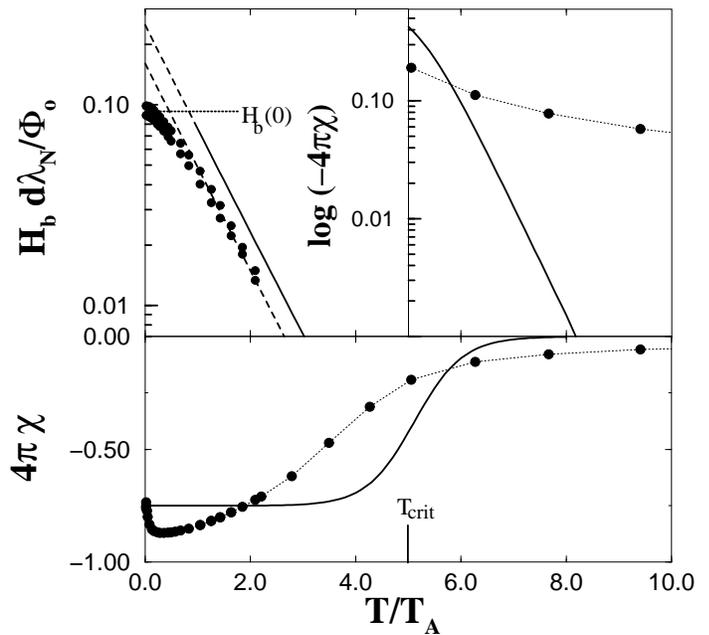,angle=-90,width=82mm}}
\caption[*]{Breakdown field $H_b\left(T\right)$ and linear 
susceptibility $4\pi\chi$ from theory and experiment (the analysis 
applies to a Ag-Nb sample of thickness $d=5.5\mu\mbox{m}$). 
Theory: results of Eqs.\ (\ref{breakdown}) and (\ref{suscept}) shown as 
solid lines; 
$H_b\left(T\right)$ is rescaled to fit the zero temperature value 
$H_b\left(0\right)$ (horizontal line) to the experiment. Experiment: data 
shown as solid dots, the dotted line is a guide to the eye\cite{motab}. 
Note that the logarithmic slope of the breakdown field is reproduced 
precisely (dashed line), while the one of the susceptibility is much 
smaller than expected.}
\label{fig3}
\end{figure}

In Fig.\ \ref{fig3} we show the two data sets for the breakdown field data 
obtained on heating and cooling a sample of thickness $d=5.5\mu\mbox{m}$
exhibiting hysteresis (the theoretical values of the super-cooled and 
super-heated fields $H_{sc}$ and $H_{sh}$ are not reached in the
experiments). The data saturates at low temperatures, in 
qualitative agreement with our theoretical analysis. Given the electron 
density in Ag, $n=5.8\cdot 10^{22}$ ($\Rightarrow \lambda_N=2.2\cdot 10^{-6}$) 
and $d=5.5 \mu \mbox{m}$, the breakdown field is determined by 
Eq.\ (\ref{b0}) and (\ref{breakdown}). Due to the idealization of our model, 
which assumes a step
function for the order parameter, as well as the difference between the planar 
and cylindrical geometry, we expect a numerical factor correcting the 
amplitude of $H_b$. Making use of the scaling factor $\approx 0.56$ in Eqs. 
(\ref{b0}) and (\ref{breakdown}), we calibrate the theoretical result to fit 
the zero temperature value $H_b\left(0\right)$, as shown in Fig.\ \ref{fig3}. 
The theoretical prediction 
for the high temperature behavior then follows from Eq.\ (\ref{breakdown}) and
is shown as a solid line in Fig.\ \ref{fig3}. Most importantly, 
Eq.\ (\ref{breakdown}) accurately reproduces the logarithmic slope $-1/T_A$ of 
the experimental data, thus correctly tracing the signature of the Andreev 
levels. The amplitude of $H_b\left(T\right)$ deviates from the data 
by the constant ratio $\approx 0.64$, which can be attributed to the presence 
of a barrier at the NS interface, see below.

An important further agreement between theory and experiment is found in the 
scaling of the breakdown field with sample thickness $d$, which was reported 
to be $\propto 1/d$, in accordance with Eq.\ (\ref{breakdown}) 
(the experimental study involved 
$10$ samples\cite{motab,bernd} with thicknesses ranging from 
$d=2.9\mu\mbox{m}$ to $d=28\mu\mbox{m}$). Similarly, the critical temperature 
determined in the experiment\cite{motapd} exhibits the same scaling 
$\propto 1/d$, in agreement with Eq.\ (\ref{tcrit}) ($T_A\propto 1/d$).

The only result in the literature on the breakdown field\cite{orsay} was 
derived from the GL equations in the dirty limit 
${\sl l}_{el} \ll \xi_D= \sqrt{v_F l_{el}/6\pi T} < d$, with the coherence 
length $\xi_D$ limited by $\lambda_D\left(d\right) \ll \xi_D < d$ 
($\lambda_D\left(x\right)$ is a space and temperature dependent penetration 
depth, see Ref.\ \cite{orsay}). The breakdown field  
\be H_D\left(T\right)\approx 1.9 \frac{\Phi_{\circ}}
{\lambda_D\left(0\right)\xi_D} 
\exp \left(-d/\xi_D\right) \label{dirtyHb} 
\ee
exhibits a simple exponential dependence on $d/\xi_D\propto \sqrt{T}$, 
the amplitude being temperature independent. Furthermore, no dependence of 
the amplitude on the thickness is present. Clearly, the experimental data 
deviates significantly from the predictions made by the GL theory.

The good agreement between the clean limit theory and experiment for the 
breakdown field does not trivially generalize to other physical 
quantities, however. In particular, the temperature 
dependence of the linear susceptibility $\chi={\cal M}/H$ exhibits distinct 
features due to the non-locality of the constitutive relation 
$j\left(a\right)$ which are not observed in the experiment.
From Eqs.\ (\ref{magnet}) and (\ref{jlinear}) we obtain the susceptibility
\be 4\pi \chi = \frac{4\pi{\cal M}}{H} =  
-\frac{3}{4}\,\frac{1}{1+3\lambda_N^2\left(T\right)/d^2}, 
\label{suscept} \ee
which exhibits a temperature dependence much like a Fermi-Dirac distribution: 
$4\pi \chi$  decays exponentially $\propto 1/\lambda^2_N\left(T\right)$ at 
large temperatures, twice as fast as the breakdown field. The susceptibility 
takes half its maximal value at $3\lambda_N^2 \left(T_{1/2}\right)/d^2 
\sim 1$, which roughly coincides with the critical temperature $T_{crit}$ 
(see Fig.\ \ref{fig3}).
Below the critical point, the susceptibility saturates as the penetration
depth decreases below the sample thickness ($\lambda_N\left(T\right) < d$). 
Due to the non-locality, the penetration depth drops out of the 
expression for $4\pi\chi \approx -3/4$ and 
we are in the regime of over-screening. The logarithmic derivative at 
$T=T_{1/2}$ is predicted to be 
$\chi'\left(T_{1/2}\right)/\chi\left(T_{1/2}\right) = 1/T_A$. 
In Fig.\ \ref{fig3} we show the linear susceptibility according to 
the clean limit predictions (\ref{suscept}) (there is no fitting parameter).
The experimental data fails to show the typical saturation of
the susceptibility expected below the critical temperature. At low temperature 
the experimental value clearly exceeds the maximal diamagnetic value $-3/4$ 
found in the clean limit (note that we do not consider the anomalous 
re-entrance effect of these samples at very low temperatures here). 
Most strikingly, the decay at large 
temperature is slower than the decay of the breakdown field, while 
Eq.\ (\ref{suscept}) predicts a decay with twice the logarithmic slope, 
see Fig.\ \ref{fig3}. Thus the magnetic behavior of the quasi-ballistic 
samples deviates from the clean limit theory, indicating that the elastic 
mean free path is of order of the thickness of the sample. We find that the 
susceptibility emerges as a very sensitive indicator of the non-locality of 
the constitutive relations.

Let us address the question whether the consideration of a insulating barrier 
at the NS interface may lift the discrepancy.
The consequences of a finite reflectivity at the NS interface on the linear 
current response has been analyzed by Higashitani and Nagai\cite{nagai}. 
Their results allow for the reflection coefficient $R$ to be  
included in the penetration depth $\lambda_N\left(T\right)$ 
by redefining the factor 
$\gamma_R\left(T_A\ll T\ll \Delta\right)=\left(1-R\right)/\left(1+R\right)$,
in Eq.\ (\ref{lambda}); $\lambda_N\left(0\right)\equiv \lambda_N$ remains 
unchanged\cite{nagai}. Inserting the modified penetration depth into 
Eq.\ (\ref{suscept}) we 
obtain the linear susceptibility. The additional factor $\gamma$ does not 
change the characteristic shape of the susceptibility (saturation, 
logarithmic slope at $T_{1/2}$, exponential decay), 
but only lowers the position of the half-value of $\chi$ to 
$T_{1/2}\approx \log\left[d\left(1-R\right)/\lambda_N\left(1+R\right)\right]$. 
Thus the finite reflection does not remedy 
the qualitative discrepancy between theory and experiment, which 
may have to be attributed to disorder or diffusive boundary scattering. 

Considering the structure of the equations we may expect the dependence on the
reflection $R$ to enter in a similar fashion into the breakdown field 
$H_b\left(T\right)$, although we note that this has not been shown rigorously. 
Eq.\ (\ref{suscept}) inserted in Eq.\ (\ref{breakdown}) gives the high 
temperature behavior, while the zero temperature result of Eq.\ (\ref{b0})
remains unchanged. We fit the breakdown field data by using first an overall 
scaling factor needed to adjust $H_b\left(0\right)$ and secondly, a finite 
reflectivity, which only enters at high temperatures. The fit of the 
high temperature behavior provides us with an estimate of the reflectivity 
$R\approx 0.21$, and is represented by the dashed line 
in Fig.\ \ref{fig3}. 

In conclusion, we have calculated the clean limit expression for the breakdown 
field separating the diamagnetic phase and the field penetration
phase by a first order transition. We have determined the spinodals, the 
critical temperature as well as the latent heat of the transition. 
In comparison with the 
experimental data on quasi-ballistic samples, we have found good agreement 
with respect to the dependences on temperature and thickness of 
$H_b\left(T,d\right)$ and $T_{crit}\left(d\right)$. The inclusion of a finite 
reflection at the NS interface permits an accurate fit of the breakdown 
field and gives an estimate for the quality of the NS interface. However, 
with regard to the linear susceptibility, the experiments disagree with the 
clean limit theory, showing the need to include additional scattering 
processes. The susceptibility thus emerges as a quantity which is very 
sensitive to the non-locality of the constitutive relations. 

We are indebted to V.\ Geshkenbein for the frequent, elucidating discussions 
throughout this work. We gratefully acknowledge A.C.\ Mota and B.\ M\"uller 
for the discussion of their experimental data. We have also benefited from a 
stimulating exchange with W.\ Belzig, C.\ Bruder, G.\ Lesovik, M.\ Sigrist, 
and A. Zaikin.

\end{multicols}


\begin{thebibliography}{99}
\bibitem{belz2}The numerical data has kindly been provided by W.
Belzig, Universit\"at Karlsruhe.
\end{thebibliography}
\end{document}